\documentclass[
]{interact}

\title{Instability cascade of strongly nonlinear gravity waves in a vertically sheared atmosphere}

\author{
    \name{G.~S. Voelker\textsuperscript{a}\thanks{CONTACT G.~S. Voelker. Email: voelker@iau.uni-frankfurt.de} and Mark Schlutow\textsuperscript{b}}
    \affil{\textsuperscript{a}Goethe University Frankfurt, Frankfurt am Main, Germany; \textsuperscript{b}Max-Planck-Institute for Biogeochemistry, Jena, Germany}
}

\usepackage[english]{babel}

\usepackage{amsmath}
\usepackage{amssymb}
\usepackage{amsfonts}
\usepackage{graphicx}
\usepackage[table]{xcolor}
\usepackage[colorlinks=true, allcolors=blue]{hyperref}
\usepackage{lineno}
\usepackage[
    authoryear,
    sort
]{natbib}

\bibpunct[, ]{[}{]}{,}{n}{,}{,}
\makeatletter
\def\NAT@def@citea{\def\@citea{\NAT@separator}}
\makeatother

\begin{document}

\articletype{DRAFT FOR SUBMISSION}  


\renewcommand{\v}[1]{\ensuremath{\boldsymbol{#1}}}
\newcommand{\vq}[1]{\v{#1}}
\newcommand{\nab}[1][]{\ensuremath{\nabla_{#1}}}
\newcommand{\nabh}[1][]{\ensuremath{\nabla_{#1\mbox{\tiny h}}}}
\newcommand{\nabhh}{\ensuremath{\hat{\nabla}_{\mbox{\tiny h}}}}

\newcommand{\vex}{\ensuremath{\v{e}_x}}
\newcommand{\vey}{\ensuremath{\v{e}_y}}
\newcommand{\vez}{\ensuremath{\v{e}_z}}
\newcommand{\vezh}{\ensuremath{\v{e}_{\hat{z}}}}

\newcommand{\dt}{\ensuremath{\partial_t}}
\newcommand{\dth}{\ensuremath{\partial_{\hat{t}}}}
\newcommand{\dT}[1][]{\ensuremath{\partial_{T_{#1}}}}
\newcommand{\dX}[1][]{\ensuremath{\partial_{X_{#1}}}}
\newcommand{\dY}[1][]{\ensuremath{\partial_{Y_{#1}}}}
\newcommand{\dZ}[1][]{\ensuremath{\partial_{Z_{#1}}}}
\newcommand{\dtau}[1][]{\ensuremath{\partial_{\tau_{#1}}}}
\newcommand{\ds}[1][]{\ensuremath{\partial_{s_{#1}}}}
\newcommand{\dm}{\ensuremath{\partial_{m}}}
\newcommand{\dk}[1][]{\ensuremath{\partial_{k_{#1}}}}

\newcommand{\dxx}{\ensuremath{\frac{\partial^2}{\partial x^2}}}
\newcommand{\dyy}{\ensuremath{\frac{\partial^2}{\partial y^2}}}
\newcommand{\dzz}{\ensuremath{\frac{\partial^2}{\partial z^2}}}
\newcommand{\dx}{\ensuremath{\partial_x}}
\newcommand{\dy}{\ensuremath{\partial_y}}
\newcommand{\dz}{\ensuremath{\partial_z}}
\newcommand{\dzh}{\ensuremath{\partial_{\hat{z}}}}

\newcommand{\Ds}{\ensuremath{d_s}}
\newcommand{\dsig}{\ensuremath{\partial_\sigma}}
\newcommand{\Dsig}{\ensuremath{d_\sigma}}

\newcommand{\Dt}{\ensuremath{D_t}}
\newcommand{\Dth}{\ensuremath{D_{\hat{t}}}}

\newcommand{\pt}[1]{\ensuremath{\frac{\partial #1}{\partial t}}}
\newcommand{\pz}[1]{\ensuremath{\frac{\partial #1}{\partial z}}}

\newcommand{\pzz}[1]{\ensuremath{\frac{\partial^2 #1}{\partial z^2}}}

\newcommand{\nn}{\nonumber}

\renewcommand{\*}{\ensuremath{\cdot}}
\renewcommand{\=}{\ensuremath{\, = \,}}

\newcommand{\eval}[1][]{\ensuremath{|_{_{#1}}}}
\def\xstrut{\rule{0pt}{2ex}}
\newcommand{\conj}{\ensuremath{\hspace{-.4em}~^{\strut *}}}
\newcommand{\res}{\ensuremath{^{\mbox{\tiny (r)}}}}
\newcommand{\ct}{\ensuremath{\hspace{-.4em}~^{\strut +}}}

\newcommand{\avg}[2][1]{\ensuremath{\overline{#2}^{~(T_{#1}, \v{X}_{#1})}}}
\newcommand{\tavg}[2][1]{\ensuremath{\overline{#2}^{~T_{#1}}}}

\newcommand{\half}{\ensuremath{\frac{1}{2}}}
\newcommand{\quart}{\ensuremath{1/4}} 


\newcommand{\tH}{\ensuremath{ \tilde{H} }}
\newcommand{\Lw}{\ensuremath{ \tilde{L}_{\mbox{\tiny w}} }}
\newcommand{\Lmf}{\ensuremath{ \tilde{L}_{\mbox{\tiny mf}} }}
\newcommand{\tB}{\ensuremath{ \tilde{B} }}
\newcommand{\tP}{\ensuremath{ \tilde{P} }}
\newcommand{\tU}{\ensuremath{ \tilde{U} }}
\newcommand{\tV}{\ensuremath{ \tilde{V} }}
\newcommand{\tW}{\ensuremath{ \tilde{W} }}

\newcommand{\tT}[1][]{\ensuremath{\widetilde{T}_{\mbox{\tiny #1}}}}
\newcommand{\tX}[1][]{\ensuremath{\widetilde{X}_{\mbox{\tiny #1}}}}
\newcommand{\tl}{\ensuremath{\tilde{l}}}
\newcommand{\tf}{\ensuremath{\tilde{f}}}
\newcommand{\tN}{\ensuremath{\tilde{N}}}

\newcommand{\Tref}{\ensuremath{ \theta_{\mbox{\tiny ref}} }}

\newcommand{\Ro}{\ensuremath{ \mathsf{Ro} }}
\newcommand{\Ri}{\ensuremath{ \mathsf{Ri} }}


\newcommand{\ep}{\epsilon}
\newcommand{\be}{\beta}
\newcommand{\ga}{\gamma}
\newcommand{\de}{\delta}
\newcommand{\degr}{\ensuremath{^{\circ}}}
\renewcommand{\deg}{\degr}

\newcommand{\fn}{\ensuremath{f_0}}

\newcommand{\phib}{\ensuremath{\phi_\beta}}
\newcommand{\phig}{\ensuremath{\phi_\gamma}}
\newcommand{\phid}{\ensuremath{\phi_\delta}}

\newcommand{\vphib}{\ensuremath{\varphi_\beta}}
\newcommand{\vphig}{\ensuremath{\varphi_\gamma}}
\newcommand{\vphid}{\ensuremath{\varphi_\delta}}

\newcommand{\omb}{\ensuremath{\omega_\beta}}
\newcommand{\omg}{\ensuremath{\omega_\gamma}}
\newcommand{\omd}{\ensuremath{\omega_\delta}}

\newcommand{\omh}[1][]{\ensuremath{\hat{\omega}_{#1}}}
\newcommand{\omhb}{\ensuremath{\hat{\omega}_\beta}}
\newcommand{\omhg}{\ensuremath{\hat{\omega}_\gamma}}
\newcommand{\omhd}{\ensuremath{\hat{\omega}_\delta}}

\newcommand{\vk}[1][]{\ensuremath{\v{k}_{#1}}}
\newcommand{\vkh}[1][]{\ensuremath{\v{k}_{h#1}}}
\newcommand{\vkb}{\ensuremath{\v{k}_\beta}}
\newcommand{\vkbh}{\ensuremath{\v{k}_{\beta h}}}
\newcommand{\vkg}{\ensuremath{\v{k}_\gamma}}
\newcommand{\vkgh}{\ensuremath{\v{k}_{\gamma h}}}
\newcommand{\vkd}{\ensuremath{\v{k}_\delta}}
\newcommand{\vkdh}{\ensuremath{\v{k}_{\delta h}}}

\newcommand{\kb}{\ensuremath{k_\beta}}
\newcommand{\kg}{\ensuremath{k_\gamma}}
\newcommand{\kd}{\ensuremath{k_\delta}}

\newcommand{\lb}{\ensuremath{l_\beta}}
\renewcommand{\lg}{\ensuremath{l_\gamma}}
\newcommand{\ld}{\ensuremath{l_\delta}}

\newcommand{\mb}{\ensuremath{m_\beta}}
\newcommand{\mg}{\ensuremath{m_\gamma}}
\newcommand{\md}{\ensuremath{m_\delta}}

\newcommand{\vv}{\v{v}}
\newcommand{\vu}{\v{u}}

\newcommand{\vun}[1][]{\ensuremath{\v{U}^{(#1)}_0}}
\newcommand{\vub}[1][]{\ensuremath{\v{U}^{(#1)}_\beta}}
\newcommand{\vug}[1][]{\ensuremath{\v{U}^{(#1)}_\gamma}}
\newcommand{\vud}[1][]{\ensuremath{\v{U}^{(#1)}_\delta}}

\newcommand{\vvn}[1][]{\ensuremath{\v{V}^{(#1)}_0}}
\newcommand{\vvb}[1][]{\ensuremath{\v{V}^{(#1)}_\beta}}
\newcommand{\vvg}[1][]{\ensuremath{\v{V}^{(#1)}_\gamma}}
\newcommand{\vvd}[1][]{\ensuremath{\v{V}^{(#1)}_\delta}}
\newcommand{\vvw}[2][1]{\ensuremath{\v{V}^{(#1)}_{#2}}}

\newcommand{\un}[1][]{\ensuremath{U^{(#1)}_0}}
\newcommand{\vn}[1][]{\ensuremath{V^{(#1)}_0}}
\newcommand{\wn}[1][]{\ensuremath{W^{(#1)}_0}}
\newcommand{\ub}[1][]{\ensuremath{U^{(#1)}_\beta}}
\newcommand{\vb}[1][]{\ensuremath{V^{(#1)}_\beta}}

\newcommand{\psig}{\ensuremath{\psi_{\mbox{\tiny g}}}}

\newcommand{\w}[1][]{\ensuremath{W^{(1)}_{#1}}}
\newcommand{\Wn}[1][]{\ensuremath{W^{(#1)}_0}}
\newcommand{\Wb}[1][]{\ensuremath{W^{(#1)}_\beta}}
\newcommand{\Wg}[1][]{\ensuremath{W^{(#1)}_\gamma}}
\newcommand{\Wd}[1][]{\ensuremath{W^{(#1)}_\delta}}

\newcommand{\bn}[1][]{\ensuremath{B^{(#1)}_0}}
\newcommand{\bb}[1][]{\ensuremath{B^{(#1)}_\beta}}
\newcommand{\bg}[1][]{\ensuremath{B^{(#1)}_\gamma}}
\newcommand{\bd}[1][]{\ensuremath{B^{(#1)}_\delta}}

\newcommand{\thbar}[1][]{\ensuremath{\bar{\Theta}^{(#1)}}}
\newcommand{\thn}[1][]{\ensuremath{\Theta^{(#1)}_0}}
\newcommand{\thb}[1][]{\ensuremath{\Theta^{(#1)}_\beta}}
\newcommand{\thg}[1][]{\ensuremath{\Theta^{(#1)}_\gamma}}
\newcommand{\thd}[1][]{\ensuremath{\Theta^{(#1)}_\delta}}

\newcommand{\pn}{\ensuremath{P_0}}
\newcommand{\pb}[1][]{\ensuremath{P^{(#1)}_\beta}}

\newcommand{\pibar}[1][]{\ensuremath{\bar{\Pi}^{(#1)}}}
\newcommand{\pin}[1][]{\ensuremath{\Pi^{(#1)}_0}}
\newcommand{\pib}[1][]{\ensuremath{\Pi^{(#1)}_\beta}}

\newcommand{\Rbar}{\ensuremath{\bar{R}^{(0)}}}

\newcommand{\Tbgd}[1][]{\ensuremath{T_{\beta\gamma\delta}^{(#1)}}}
\newcommand{\Tb}[1][]{\ensuremath{T_{\beta}^{(#1)}}}
\newcommand{\Mb}{\ensuremath{M_\beta}}
\newcommand{\Zb}[1][]{\ensuremath{Z_\beta^{(#1)}}}
\newcommand{\Zbp}[1][]{\ensuremath{Z_\beta^{(#1)+}}}
\newcommand{\zb}[1][]{\ensuremath{\zeta_\beta^{(#1)}}}
\newcommand{\zg}[1][]{\ensuremath{\zeta_\gamma^{(#1)}}}
\newcommand{\zd}[1][]{\ensuremath{\zeta_\delta^{(#1)}}}
\newcommand{\zw}[2][1]{\ensuremath{\zeta_{#2}^{(#1)}}}
\newcommand{\zn}{\ensuremath{\zeta_0}}
\newcommand{\Rb}{\ensuremath{R_\beta}}

\newcommand{\xibp}[1][1]{\ensuremath{\xi_\beta^{(#1)+}}}
\newcommand{\xib}[1][1]{\ensuremath{\xi_\beta^{(#1)}}}
\newcommand{\vcghb}{\ensuremath{\hat{\v{c}}_{g\beta}}}
\newcommand{\vcghbh}{\ensuremath{\hat{\v{c}}_{g\beta h}}}
\newcommand{\vcgb}{\ensuremath{\v{c}_{g\beta}}}
\newcommand{\vcgg}{\ensuremath{\v{c}_{g\gamma}}}
\newcommand{\vcgd}{\ensuremath{\v{c}_{g\delta}}}
\newcommand{\vcg}[1][]{\ensuremath{\v{c}_{g #1}}}
\newcommand{\cgzb}{\ensuremath{c_{g\beta z}}}
\newcommand{\cgz}[1][]{\ensuremath{c_{g z #1}}}

\newcommand{\cgxhb}{\ensuremath{\hat{c}_{g\beta x}}}
\newcommand{\cgyhb}{\ensuremath{\hat{c}_{g\beta y}}}
\newcommand{\cgzhb}{\ensuremath{\hat{c}_{g\beta z}}}

\newcommand{\cgzg}{\ensuremath{c_{g\gamma z}}}
\newcommand{\cgzd}{\ensuremath{c_{g\delta z}}}
\newcommand{\cgzs}{\ensuremath{c_{gz}}}

\newcommand{\Eb}[1][1]{\ensuremath{E_\beta^{(#1)}}}
\newcommand{\En}{\ensuremath{E_0}}
\newcommand{\E}[1][]{\ensuremath{E_{#1}}}
\newcommand{\Ab}[1][1]{\ensuremath{\A_\beta^{(#1)}}}
\newcommand{\vpbh}[1][1]{\ensuremath{\v{p}^{(#1)}_{\be h}}}
\newcommand{\vpb}[1][1]{\ensuremath{\v{p}^{(#1)}_{\be}}}

\newcommand{\Pn}{\ensuremath{\pi_0}}
\newcommand{\PVn}{\ensuremath{\Pi_0}}

\newcommand{\Ap}[1][]{\ensuremath{A_{#1}^+}}
\newcommand{\Am}[1][]{\ensuremath{A_{#1}^-}}
\newcommand{\Apm}[1][]{\ensuremath{A_{#1}^\pm}}
\newcommand{\Cp}[1][]{\ensuremath{C_{#1}^+}}
\newcommand{\Cm}[1][]{\ensuremath{C_{#1}^-}}
\newcommand{\Cpm}[1][]{\ensuremath{C_{#1}^\pm}}
\newcommand{\Abgd}[1][]{\ensuremath{A_{\be\ga\de}^{#1}}}
\newcommand{\Bbgd}[1][]{\ensuremath{B_{\be\ga\de}^{#1}}}
\newcommand{\coeff}[2][123]{\ensuremath{A_{#1}^{#2}}}
\newcommand{\varcoeff}[2][123]{\ensuremath{B_{#1}^{#2}}}

\newcommand{\chit}{\ensuremath{\chi_T}}
\newcommand{\chiz}{\ensuremath{\chi_Z}}
\newcommand{\vchix}{\ensuremath{\v{\chi}_X}}

\newcommand{\cp}{\ensuremath{c_{\mbox{\tiny p}}}}
\newcommand{\cv}{\ensuremath{c_{\mbox{\tiny V}}}}

\newcommand{\Lh}{\ensuremath{L_{\mbox{\tiny h}}}}

\newcommand{\unit}[1][]{\ensuremath{\, \mathrm{#1}}}


\newcommand{\A}{\ensuremath{\mathcal A}}
\newcommand{\B}{\ensuremath{\mathcal B}}
\newcommand{\C}{\ensuremath{\mathcal C}}
\newcommand{\mathcalD}{\ensuremath{\mathcal D}}
\newcommand{\F}{\ensuremath{\mathcal F}}
\newcommand{\G}{\ensuremath{\mathcal G}}
\renewcommand{\H}{\ensuremath{\mathcal H}}
\newcommand{\J}{\ensuremath{\mathcal J}}
\renewcommand{\L}{\ensuremath{\mathcal L}}
\newcommand{\K}{\ensuremath{\mathcal K}}
\newcommand{\M}{\ensuremath{\mathcal M}}
\newcommand{\N}{\ensuremath{\mathcal N}}
\newcommand{\NN}{\ensuremath{\mathbb{N}}}
\newcommand{\NNn}{\ensuremath{\mathbb{N}_0}}
\renewcommand{\O}{\ensuremath{\mathcal O}}
\renewcommand{\P}{\ensuremath{\mathcal P}}
\newcommand{\Q}{\ensuremath{\mathcal Q}}
\newcommand{\R}{\ensuremath{\mathsf R}} 
\newcommand{\RR}{\ensuremath{\mathbb R}}
\renewcommand{\S}{\ensuremath{\mathcal S}}
\newcommand{\T}{\ensuremath{\mathcal T}}
\newcommand{\U}{\ensuremath{\mathcal U}}
\newcommand{\V}{\ensuremath{\mathcal V}}
\newcommand{\WW}{\ensuremath{\mathcal W}}
\newcommand{\X}{\ensuremath{\mathcal X}}
\newcommand{\Y}{\ensuremath{\mathcal Y}}
\newcommand{\Z}{\ensuremath{\mathcal Z}}

\newcommand{\qmq}[1]{\ensuremath{\quad\quad\mbox{#1}\quad\quad}}
\newcommand{\qand}{\ensuremath{\quad\quad\mbox{and}\quad\quad}}
\newcommand{\qwith}{\ensuremath{\quad\quad\mbox{with}\quad\quad}}


\newcommand{\req}[1]{(\ref{#1})}
\renewcommand{\eqref}[1]{Eq. (\ref{#1})}
\newcommand{\eqrefs}[2]{Eqs. (\ref{#1}) and (\ref{#2})}
\newcommand{\eqrefr}[2]{Eqs. (\ref{#1}) to (\ref{#2})}
\newcommand{\peqref}[1]{(Eq. \ref{#1})}
\newcommand{\peqrefs}[2]{(Eqs. \ref{#1} and \ref{#2})}
\newcommand{\peqrefr}[2]{(Eqs. \ref{#1} to \ref{#2})}


\newcommand{\lhs}{left-hand side }
\newcommand{\rhs}{right-hand side }
\newcommand{\ie}{i.e.~}
\newcommand{\eg}{e.g.~}
\newcommand{\cf}{cf.~}

\newcommand{\freq}{frequency }
\newcommand{\freqn}{frequency}
\newcommand{\freqs}{frequencies }
\newcommand{\frqesn}{frequencies}
\newcommand{\BVF}{Brunt-V\"ais\"al\"a frequency }

\maketitle

\begin{abstract}
Although internal gravity waves are generally recognized as an important mechanism to distribute energy through the atmosphere, their dynamics near the instability is only partially understood to date. Many types of instabilities, notably the classical modulational instability, a novel point spectrum modulational instability, the triadic resonant instability, the shear instability and the static instability have been studied mostly in idealized settings and mostly isolated from one another. Here, we identify the instability cascade of a quasi one-dimensional and stationary internal gravity wave modulated by a vertically sheared mean flow. We find indicators of various interdependent instability mechanisms which partly compete for dominance and partly follow one another. A key finding is that the particular dynamics of the local cascade depends on the sign of the background shear.
%
%
\end{abstract}

\begin{abbreviations}
    internal gravity wave (IGW),
    triadic resonant instability (TRI),
    point spectrum modulational instability (PSMI),
\end{abbreviations}



\section{Introduction}

It has been widely recognized that gravity waves affect the global circulation of Earth's atmosphere. Usually excited in the troposphere and stratosphere, gravity waves carry energy vertically as well as laterally, they drag the mean flow and lead to mixing of trace gases \citep{Bretherton1969, Fritts2003, Schlutow2014}. They are associated with the anomalous summer temperature minimum in the mesopause, the Quasi-Biannual Oscillation and the residual mean-flow circulation \citep{Lindzen1981, Becker2012, Baldwin2001} 
%
Gravity waves are ubiquitous \citep{Eckermann1999}. A pivotal role in understanding the interaction of the waves with the mean flow is wave dissipation, i.e. the dynamics of a wave that becomes unstable, overturns, breaks and vanishes into chaotic turbulence eventually.
%
This work is concerned with the very first step of this chain: the instability onset. Several instability mechanisms were identified that cause an infinitesimal perturbation of a wave to grow exponentially with time. One of the problems with the theories governing the instability mechanisms lies in the fact that they only predict instability growth rates for specific instability mechanisms in an idealized setting. In nature, however, waves are not isolated and instabilities may coincide as they differ in location or in scale. For instance, convective and shear instabilities appear on comparable spatio-temporal scales but at distinct positions relative to the wave's period. Convective instabilities grow where the buoyancy is at its minimum and the shear instability can be found where the wind shear has its maximum \citep{Lelong1998a, Lelong1998}. In contrast to convective and shear, modulational instabilities occur on scales that are much larger than the period \citep{Benjamin1967}. They are excited on scales that are comparable with the typical variation of the mean flow, the synoptic scale or the mesoscale. Another class of instability that may be found on comparable scales to the wave itself is Triadic Resonant Instability and in particular the parametric subharmonic instability \citep{Dong1991, Dauxois2018}. Not only do these instability mechanisms coincide, they may also trigger each other. Modulational instabilities, for example, are able to amplify waves locally which causes an increased buoyancy amplitude which causes convective instability. Consequently, waves can undergo an entire cascade of instabilities before they dissipate and, conversely, it is usually impossible to identify one single instability mechanism responsible for the dissipation of a wave. We argue that only if one understands the instability cascade, one is able to predict gravity wave dissipation.

Naturally, waves become unstable when their amplitude is large. In contrast to internal gravity waves in the ocean, atmospheric gravity waves tend to have as large wind amplitudes as the background or mean-flow wind, respectively. This is an indirect effect of air's compressibility that causes an exponentially decreasing ambient density with altitude. Due to energy conservation, atmospheric gravity waves gain in amplitude when they propagate upwards. Therefore, amplitudes become easily so large that neither linear nor weakly nonlinear theories are applicable. The former requires infinitesimal small and the latter finite but still small amplitudes in comparison with the mean flow \citep{Schlutow2019}. Strongly nonlinear waves with large amplitudes are, in conclusion, rather the rule than the exception in the atmosphere and therefore we want to focus, in particular, on this class of waves in this study.

Theoretical studies on gravity wave instabilities often assume homogeneous background atmosphere, i.e. constant stratification and constant ambient wind. A more realistic (but still highly idealized) scenario arises when sheared ambient wind is considered adding a substantial layer of complexity to the problem.

\citet{Schlutow2020} (hereafter SV20) carried out a theoretical investigation of the modulation equations, which describe the temporal evolution of the wave parameters, such as wave number and amplitude, in an ideal atmosphere with a very thin shear layer. This scenario is a common occurrence in the actual atmosphere, such as when a mountain wave encounters the tropospheric jet \citep{Ehard2017}. SV20 found a novel instability that is generally similar to the canonical modulation instability \citep{Sutherland2001, Schlutow2019} in unsheared backgrounds. The difference lies in the operator representing the linearized equations. Whereas the canonical modulation instability comes from the essential (continuous) spectrum of the operator, the novel type of instability is generated by the point (matrix-like) spectrum or in  other words the set of discrete eigenvalues of the operator. In particular, the latter occurs only at a lower edge of a spatially slowly varying jet and at sufficiently large amplitudes of the stationary wave.

This novel instability type is thus an addition to the many known instability mechanisms \citep{Fritts2003}. It remains, however, largely unclear how these various mechanisms interdepend and are thus linked to one another. With all these complications in mind, the aim of this study is to examine the instability cascade---from the first unstable perturbation to the knock-out mode---of strongly nonlinear gravity waves that interact with a layer of sheared background wind. 

This paper is structured as follows. In Section~\ref{sec:theory} we state the problem of the stationary refracted wave in a sheared background flow in terms of the Bretherton-Grimschaw modulation equations  \citep{Bretherton1966, Grimshaw1974}. A theoretical account on the manifold instability mechanisms is given; we review the classical (essential spectrum) modulation instability as well as the novel point-spectrum instability, then static, and shear instabilities. A brief discussion on triadic resonant instabilities concludes this section. The numerical model, that we utilize to simulate the refracted wave, is described in Section~\ref{sec:model}. Our simulation results together with analyses with respect to the various instabilities, that we found, are shown in Section~\ref{sec:observations}. The concluding Section~\ref{sec:conclusion} summarizes our results and gives some final thoughts.




\section{Instabilities of the stationary refracted wave solution}
\label{sec:theory}

In general an internal wave mode may encounter a wide range of known and individually studied instabilities. To highlight the most important instability types of a Boussinesq nonlinear internal gravity wave in a sheared background we consider the Bretherton-Grimschaw modulation equations in two dimensions as follows \citep{Achatz2010, Schlutow2017, Schlutow2020}

\begin{align}
\label{eq:govern}
    0 &= \dt k_z + \dz \omega, \nn\\
    0 &= \rho\dt a + \dz\left(\cgz\rho a\right), \\
    0 &= \rho \dt u + \dz \left(\cgz k_x \rho a \right). \nn
\end{align}

While the wave action, $a$, the wave vector, $(k_x, k_z)$, the intrinsic frequency, $\omh$, and the vertical group velocity, $\cgz = \dk[z]\omh$, are wave parameters, $u$ represents the horizontal mean flow velocity including a background flow and a wave induced part. Note that the Doppler shifted extrinsic frequency is denoted by $\omega = \omh + k_xu$. The background density, $\rho$, and the buoyancy frequency, $N$, are assumed constant for simplicity. The mean-flow velocity then decomposes as follows

\begin{align}
    u(z, t) &= k_x a(z, t) + U(z)\\
        &= k_x a(z, t) + U_1 + \frac{U_2 - U_1}{2}\tanh{\frac{z - z_0}{h}}.
\end{align}

Where we have imposed a background flow with a shape of a \textit{tangens hyperbolicus} changing continuously from a lower level with $U\approx U_1$ to an upper layer with $U\approx U_2$ on a length scale $h$. To analyze the stability of such a refracted wave SV20 make use of an asymptotic WKBJ approach with a spatially slowly varying wave amplitude and mean flow. Since the vertical wavelength and the transition length scale, $h$, are similar in magnitude the transition between $U_1$ and $U_2$ appears as a jump in the background velocity on the slowly varying scale. Additionally, we assume a stationary primary wave. Such a solution then yields vertical wavenumbers given by $K_{z,j} = -(N^2 / U_j^2 - K_x^2)^{\half}$ with $j\in\{1,\,2\}$ and a boundary condition between the two layers, $A_2 \cgz[,2]= A_1 \cgz[,1]$. Here we define upper case variables as explicit solutions for the steady state primary wave. For convenience one may also define the relative frequency square, $J_j = N^2 / K_x^2U_j^2$, and the relative wave amplitude, $\alpha = |\B K_z| / N^2$, where $\B$ denotes the buoyancy amplitude of the wave. Note that in such a notation $\alpha = 1$ marks the threshold for static instability from linear theory \cite[\eg][]{Sutherland2010, Achatz2022}. In the theoretical analysis we thus assume a stable incident wave with $\alpha < 1$ everywhere.\\
Demanding a transient wave solution, one finds the condition for non-evanescence 
\begin{align}
    J_j = \frac{N^2}{K_x^2U_j^2} > 1,
\end{align}
or equivalently $|\omh| < N$. Performing an instability analysis on the above solution one may find a description of the classical modulational instability as well as the point spectrum modulational instability (PSMI). Additionally but without reproducing the theory here we consider the triadic resonant instability, the shear instability, and the static instability mechanisms.

\subsection{Classical Modulational Instability}
\label{susec:class_mod_insta}

Following \citet{Kapitula2013,Schlutow2019} and SV20 one may use a perturbation ansatz to find both the continuous (essential) spectrum and the point (matrix-like) spectrum of the operator resulting from linearizing \eqref{eq:govern}. The former, then, provides us with the classical modulational stability criterion 

\begin{align}
    \label{eq:modulational-instability}
    J_j > \frac{3}{2}.
\end{align}

\noindent

Rather than adding another study to the often considered classical modulational instability \citep{Sutherland2001, Fritts2015} we aim at including the PSMI by choosing the initial conditions accordingly. The latter is ensured through demanding the wavenumber aspect ratio $1 / \sqrt{2} < |K_{z}| / K_x$.

\subsection{Point spectrum modulational instability}
\label{subsec:point-spectrum-inst}

In addition, the point spectrum yields another type of modulational instability. It predicts a statically stable primary wave to be unstable due to the point spectrum but stable according to the essential spectrum if
\begin{align}
    \label{eq:stability-point-spectrum}
    1 &> \alpha_1^2 > \frac{2}{J_1}\frac{(J_1 - 1)^2}{2J_1 - 3} \qand |U_1|<|U_2|.
\end{align}
Hence, $\alpha_1$ and $\alpha_2$ correspond to the amplitudes outside and inside the jet , respectively and follow the relationship $\alpha_1 > \alpha_2$. It is furthermore worth mentioning that the instability criterion \peqref{eq:stability-point-spectrum} requires a normalized wave amplitude as large as $\alpha_1 > \sqrt{8/9}$ for the instability to occur. One may argue that such large amplitudes are rare events. With respect to the atmosphere, where the present Boussinesq analysis may be regarded as a local approximation, the anelastic amplification however makes large amplitudes a common phenomenon. For more details we refer the interested reader to SV20 and the references therein.

\subsection{Triadic Resonant Instability}
\label{subsec:triadic-insta}

Triadic resonant instabilities (TRI) are considered one of the most important mechanism for spectral (and non-local) energy transfer between internal gravity waves. TRI general occurs when there are three spectral wave components which approximately fulfill the resonance conditions \citep{McComas1977, Yeh1985, Dong1991}
\begin{align}
    \label{eq:resonance-conditions}
    \begin{split}
        \vk[1] &= \vk[2] + \vk[3],\\
        \omh[1] &= \omh[2] + \omh[3]. 
    \end{split}
\end{align}
It is worth mentioning that the instability may also grow when one of the three spectral components has a zero amplitude, thus generating a third triad member. It is however generally not applicable for an incidentally monochromatic wave. In such a case---as considered here---only a perturbation with a corresponding spectral component may trigger the instability. Such a perturbation could for instance be generated through another instability mechanism and subsequently act as a triad member of a resonant or near-resonant triad. Most attention has been given to perturbations which are very close to the initial wave component leading to interactions commonly classified as parametric subharmonic instability \citep{McComas1977, Yeh1985}. Considering a stationary incident wave the resonance conditions \peqref{eq:resonance-conditions} predict the growth of a vertically propagating wave mode with a horizontal wavenumber $k'_x \approx 2 K_x$. Following recent insights into modulated TRI modulation through wind shear may reduce the growth rate of a generated mode through effectively narrowing the spectral interaction window for near-resonant interactions \citep{Voelker2020}.

\subsection{Static and Shear Instabilities}
\label{subsec:stat-shear-insta}

Gravity waves may become unstable with respect to shear instability when the local Richardson number $\mathrm{Ri}$ falls below a quarter \citep[Kelvin-Helmholtz instability;][]{Miles1961, Howard1961}. The Richardson number is canonically defined as 
\begin{align}
    \mathrm{Ri}=\frac{N^2+g\theta_0^{-1}\partial_z\theta'}{\left[\partial_z (u'+u)\right]^2}
\end{align}
which comprises the ratio of stratification to wind shear. An introduction on shear instabilities of gravity waves is given in \citep[][p 142]{Nappo2013}.

Note that the Richardson criterion is only a necessary condition for shear instability and only applies for horizontally parallel flows. Gravity waves are shear waves, i.e. the velocity field is sheared along the direction of propagation. Strictly speaking, the assumption of horizontally parallel flow is only valid for hydrostatic gravity waves. For strongly nonlinear, non-hydrostatic gravity waves, however, the critical Richardson number may be modified \citep{Liu2010}.  

The local Richardson number may be written in terms of the phase $\phi=k_xx+k_zz$ as
\begin{align}
    \label{eq:local_rich}
    \mathrm{Ri}=\frac{|\vk|^2}{k_z^2}\frac{1-\alpha\sin(\phi)}{\alpha^2\cos^2(\phi)}
\end{align}
according to \citet{Lelong1998a}.

Studying \eqref{eq:local_rich}, we learn that the Richardson criterion depends on the phase and the relative wave amplitude. Only when $\alpha$ exceeds unity, the Richardson criterion, modified or not, can be fulfilled. In other words, a shear unstable wave is also unstable with respect to static instability. With regard to the phase, the wave becomes statically unstable where $\phi=\pi/2$ or where the buoyancy field of the perturbation has its maximum. The Kelvin-Helmholtz instability appears consequently where the shear is maximized at $\phi=0$. As a final remark of this section we want to point out that the initial conditions of our simulations are neither statically nor dynamically unstable as we assume initially $\alpha<1$.

\section{Model description}
\label{sec:model}

With the above described instability mechanisms in mind we consider a nonlinear stationary wave in a sheared background flow. Such a solution is in principle stable under the assumptions of linear theory but will prove to exhibit a range of growing modes forming a cascade of instabilities which ultimately leads to the breakdown of the stationary parent wave. Here, we perform the analysis using the Large Eddy Simulation code \textit{PincFlow} with a second order MUSCL scheme and a MC flux limiter \citep{Rieper2013,Wilhelm2018,Schmid2021}. To accommodate the assumption of an incompressible flow we utilize the Boussinesq mode and integrate with an explicit third-order Runge-Kutta scheme \citep{Williamson1980}.

\begin{table}
    \centering
    \caption{Summary of relevant model parameters.}
    \rowcolors{2}{gray!20}{white}
    \begin{tabular}{llrl}
        \rowcolor{gray!40}
        parameter & description & \multicolumn{2}{c}{value} \\
        $\Delta x$ & hor. grid spacing & $161.3$ & $\unit{m}$\\
        $\Delta z$ & vert. grid spacing & $39.0$ & $\unit{m}$ \\
        $z_{\mbox{\tiny max}}$ & vert. domain extent & $79924.34$ & $\unit{m}$ \\
        $x_{\mbox{\tiny max}}$ & hor. domain extent & $5000$ & $\unit{m}$ \\
        $N$ & buoyancy frequency & $0.01$ & $\unit{s^{-1}}$\\
        $\lambda_x = 2\pi / K_x$ & hor. wavelength & $5000$ & $\unit{m}$\\
        $z_1$ & lower jet edge & $20,000$ & $\unit{m}$\\
        $z_2$ & upper jet edge & $60,000$ & $\unit{m}$\\
        $U_1$ & wind outside jet & $-4.5$ & $\unit{m\,s^{-1}}$\\
        $U_2$ & wind within jet & $-5.5$ & $\unit{m\,s^{-1}}$\\
        $h$ & transition height & $1000$ & $\unit{m}$
    \end{tabular}
    \label{tbl:model-parameters}
\end{table}

Given the conditions by the above theoretical considerations we choose the parameters summarized in Tbl. \ref{tbl:model-parameters}. To accommodate a modulated stationary initial mode we set up the model with periodic boundary conditions and a jet embedded in a background flow as follows
\begin{align}
    U(z) &= 
    \begin{cases}
        U_1,  & z < z_1 - 5h,\\
        U_1 + \frac{U_2 - U_1}{2}\tanh{\frac{z - z_1}{h}} & z \in \left[ z_1 - 5h, z_1 + 5h \right),\\
        U_2,  & z \in \left[ z_1 + 5h, z_2 - 5h \right),\\
        U_2 + \frac{U_1 - U_2}{2}\tanh{\frac{z - z_2}{h}} & z \in \left[ z_2 - 5h, z_2 + 5h \right),\\
        U_1,  & z \geq z_2 + 5h.
    \end{cases}
\end{align}
We thus embed a jet with velocity $U_2$ within a background with velocity $U_1$ with smooth edges on the top and the bottom. Correspondingly, $z_1$, and $z_2$ denote the lower and upper jet edges, and $h$ is the transition scale as before. The vertical wavenumber, $K_z$, is then
\begin{align}
K_{z,j}(z) = -\sqrt{\frac{N^2}{U(z)^2}- K_x^2},
\end{align}
such that the wavenumber ratio is fulfills $|K_{z}| / K_x \in (1.05, 1.46)$ and does not permit modulational instability in the initial conditions. Moreover we ensure a conserved wave action flux by setting $c_{gz}A = \text{const.}$ and choosing the amplitude $\alpha$ accordingly. Finally, we integrate the phase of the wave and choose a domain size, $(x_{max}, z_{max})$, such that the wave phase is continuous at the periodic vertical boundaries. The Brunt-V\"ais\"al\"a frequency, $N$, is constant. In this setup we can thus compare the behavior of the wave at both the lower as well as the upper jet edge. To ensure that the two edge regions evolve approximately independently, the heights of the jet edges are set such that they are well separated from each other and the periodic domain boundaries by approximately $40h$.

Seeking instability mechanisms in numerical experiments it is important to consider that numerical errors can trigger instability mechanisms in marginally stable initial conditions. To avoid this effect we rely on the variational diminishing discretization of \textit{PincFlow} which inhibits spurious oscillations to propagate through the numerical solution.

\section{Observed instabilities}
\label{sec:observations}

\begin{figure}
    \includegraphics[width=1\textwidth]{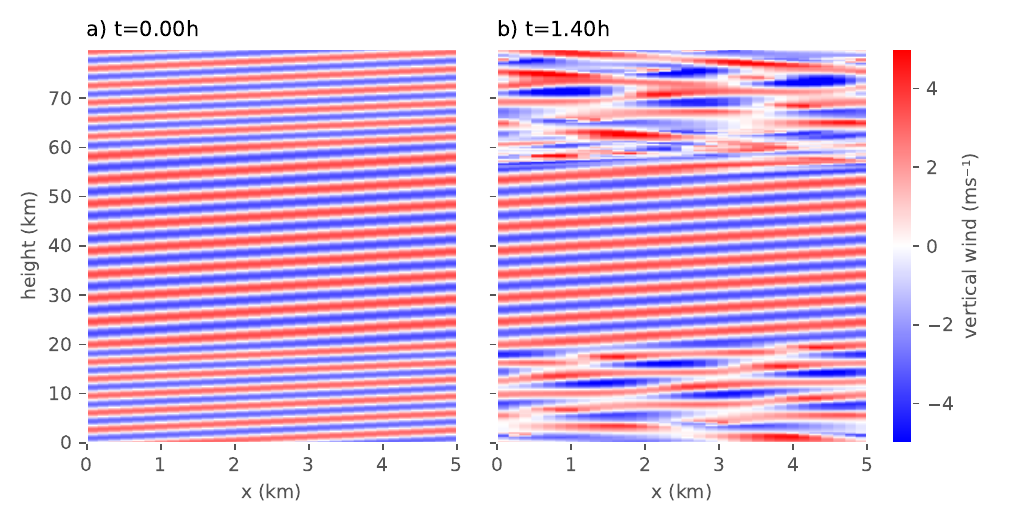}
    \caption{Snapshots of the vertical wind in the full domain of the simulation at the initial conditions (a) and after $1.4\unit{h}$ (b) simulated time.}
    \label{fig:snapshots}
\end{figure}

\begin{figure}
    \centering
    \includegraphics[width=.5\columnwidth]{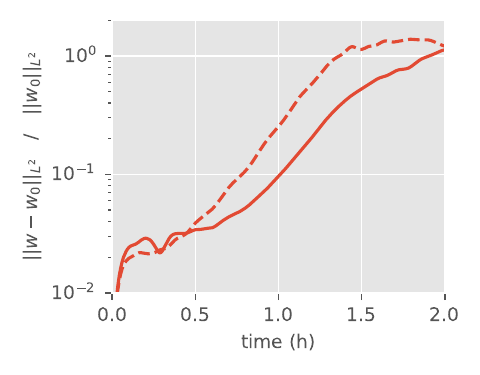}
    \caption{Normalized L$^2$-norm of $w$ integrated over the intervals $[z_1 - 5h, z_1 + 5h]$ (solid line) and $[z_2 - 5h, z_2 + 5h]$ (dashed line). The perturbation growth rates associated to the two jet edges show distinct values suggesting different growth mechanisms to be dominant there.}
    \label{fig:l2-norm}
\end{figure}

Using the model setup described above one may expect instabilities to develop and break the very large amplitude incident wave. It is that initial phase of the simulation in which the instabilities develop that we are particularly interested in. For illustration we show snapshots of the vertical velocities in the full domain for the initial conditions and after the instabilities have developed (Fig. \ref{fig:snapshots}). One possible way to identify that deviation from  the initial state is quantifying the $L_2$-norm of the vertical perturbations (\cf SV20). Integrating over the domains near the upper and lower jet edges we find that the perturbations do indeed grow exponentially but show distinct growth rates until they are reaching a saturation state at around $1.5-2\unit{h}$ (Fig. \ref{fig:l2-norm}). What is more, we find that instabilities at the upper jet grow faster and earlier compared to the lower edge. After approximately $2\unit{h}$ the initial wave is broken and the solution is transitioning to turbulent behavior. 

\begin{figure*}
    \includegraphics[width=1\textwidth]{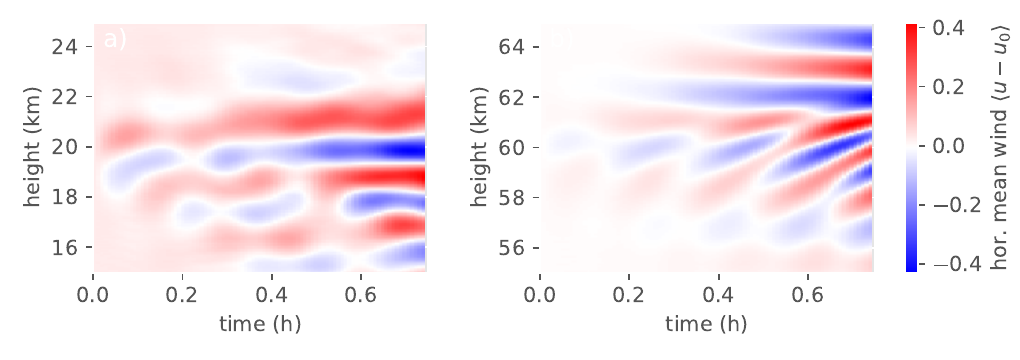}
    \caption{Hovmoeller plots of the horizontally averaged horizontal wind deviations, $\left\langle u(x, z, t) - u(x, z, 0)\right\rangle_x$, at the lower jet edge (a) and the upper jet edge (b). The two regimes show distinct generation of stationary (a and b) and transient structures (b).}
    \label{fig:mean-wind}
\end{figure*}

As for the spatial structure of the perturbation-induced mean-wind we find mostly stationary structures at the lower edge and strong transient features at the upper edge (Fig. \ref{fig:mean-wind}). This suggests that different instability mechanisms or combinations of instabilities govern the two spatial regimes of interest. To further illustrate the two distinct instability cascades we analyze the simulation results with respect to various indicators of the previously mentioned instability mechanisms.

\subsection{Point spectrum modulational instability}

\begin{figure*}
    \includegraphics[width=1\textwidth]{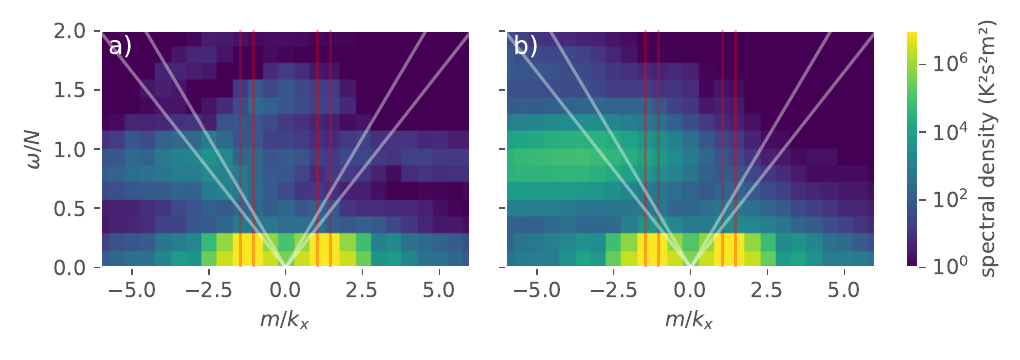}
    \caption{Wavenumber-frequency spectra of the potential temperature deviations near the lower jet edge (a) and the upper jet edge (b). The spectra are computed using Hanning windows in time and the vertical axis. The resulting spectra are then horizontally averaged. The red vertical lines indicate the vertical wavenumbers of the primary wave inside and outside the jet. Additionally, the white functions show the theoretical spectral range of the PSMI.}
    \label{fig:wn-freq-spectra}
\end{figure*}

First, we would like to highlight that the non-dimensional amplitude $\alpha=0.975$ satisfies the condition of the predicted PSMI (SV20) and may thus be observed at the lower jet edge. Also, we would like to remind the reader that the instability is predicted to be vertically localized exhibiting its maximum amplitude at the jet edge with a horizontal wavelength equal to the primary wave (\cf section \ref{subsec:point-spectrum-inst} and SV20). Thus, such instabilities would be expected to be visible in the horizontal mean wind (Fig. \ref{fig:mean-wind}) and the periodogram for the horizontal mode-1 (Fig. \ref{fig:periodogram}a and c, also see Sec. \ref{subsec:triadic-insta}). Indeed, upon visual inspection of the perturbation-induced mean wind one may find localized and temporally stationary structures at the lower jet edge (Fig. \ref{fig:mean-wind}, left panel). Similarly, the horizontal mode-1 perturbation signal reveals dominantly stationary structures similar to the expected instability at tht lower jet edge (Fig. \ref{fig:periodogram}c). However, the induced mean-wind's properties are not easily identified and associated with the PSMI. 

Moreover, the theory according to SV20 predicts a linear relationship between the temporal growth rate, $\lambda$, and the exponential spatial decay rate, $\sigma$, with a proportionality coefficient given by the wave properties and the background wind. It should be noted that the theory permits both the growth and the spatial decay rates to have an imaginary component that would correspond to oscillatory behavior. We thus compute wavenumber-frequency spectra and insert the range of expected growth rates (white lines, Fig. \ref{fig:wn-freq-spectra})). The dominant signal is, as one might expect, the primary wave at the corresponding wavenumbers and zero frequency. Moreover, we observe that the spectra are asymmetric with respect to the sign of the wavenumber of the growing perturbations. While these perturbations exhibit spectral energy within the expected range for the PSMI it may be difficult to interpret the spectra for various reasons. Firstly, we expect this type of instability to only exist at the lower jet edge (Fig. \ref{fig:wn-freq-spectra}a). However, the upper jet edge (Fig. \ref{fig:wn-freq-spectra}b) exhibits energy at similar wavenumber and frequency ranges, albeit in a more broad spectral region. Secondly, the spectral energy within the expected range does not dominate the perturbation spectrum. The PSMI may thus occur, if present, in combination with other instabilities. Ultimately, we may not unambiguously identify this novel type of instability but conclude that the present numerical experiments permit the instability to be embedded in the cascade of instabilities leading to the breakdown of the primary wave.

\subsection{Triadic resonant interaction}

\begin{figure*}
    \includegraphics[width=1\textwidth]{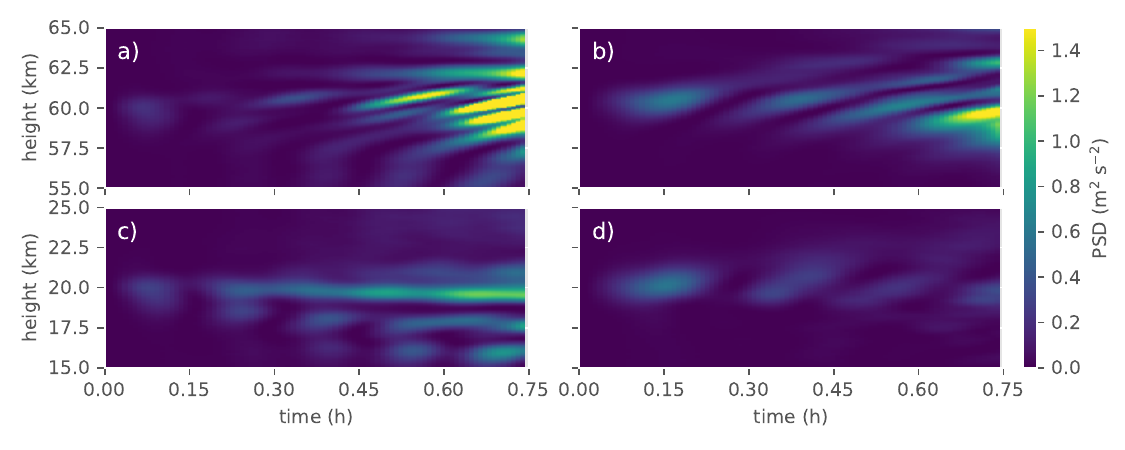}
    \caption{Power spectral densities for horizontal modes with $k_x=K_x$ (a, c) and $k_x=2K_x$ (b, d) from horizontal periodograms of  $u(x, z, t) - u(x, z, 0)$. While the upper jet edge is dominated by transient mode-1 and mode-2 wave components, the lower jet edge mostly shows a stationary mode-1 perturbation.}
    \label{fig:periodogram}
\end{figure*}

Given the quasi monochromatic incident wave and the resonance conditions \peqref{eq:resonance-conditions} we expect the TRI to generate vertically propagating wave components with a horizontal wavenumber $k'_x \approx 2 K_x$. Utilizing the horizontally periodic boundary conditions we employ horizontal periodograms of the perturbation signal, $u - u(t=0)$, to identify the growing spectral components (\cf Fig. \ref{fig:periodogram}). We find that at the upper jet edge dominantly transient components with horizontal wavenumbers $K_x$ and $2K_x$ are generated. At the lower jet edge, however, we find that a dominantly stationary component with horizontal wavenumber $K_x$ is generated. At the same time the generation of transient mode-2 waves seems greatly reduced with respect to the upper jet edge.

In general, $2K_x$ components may also be introduced through the generation of higher harmonics associated to the primary wave \citep[\eg][]{Achatz2017}. These higher harmonics, being coupled to a stationary primary wave with a zero extrinsic frequency, $\omega=0$, would be set to have a vanishing extrinsic frequency and stationary phases as well. Finding mostly transient wave structures in the horizontal mode-2 components we conclude that the TRI may be the dominant process generating aforementioned mode-2 components therein.

\subsection{Shear and static instabilities}

Albeit being stable initially with respect to shear instability it may develop as a part of the cascade of instabilities during the break down of the incident gravity wave. That is, modulation of the stationary parent wave may induce shear instabilities through changes in the vertical wavenumbers. Also, growing perturbations may become non-linear and eventually exhibit shear instabilities themselves. As noted in Sec. \ref{subsec:stat-shear-insta}, the necessary condition for shear instability to occur is fulfilled where the local Richardson number falls below one quarter, $\mathrm{Ri}\leq \quart$ \peqref{eq:local_rich}.

\begin{figure*}
    \includegraphics[width=1\textwidth]{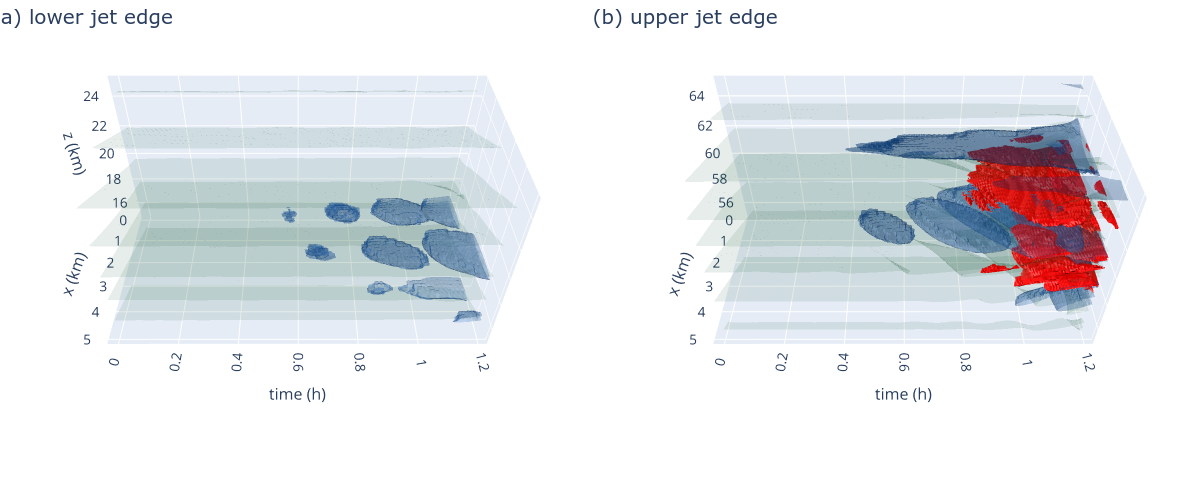}
    \caption{
        Hovmoeller diagrams of zero total vorticity and $\Ri=\quart$ for the lower jet edge (a) and the upper jet edge (b). Blue and red isosurfaces correspond to winds and potential temperatures associated to the horizontal mode 1 and higher modes, respectively. Green isosurfaces show manifolds of zero total vorticity.
    }
    \label{fig:vorticity_up}
\end{figure*}

In particular, we observe that well after the onset of first instabilities local areas develop flow characterized by a Richardson number smaller \quart~(Figs. \ref{fig:l2-norm} and \ref{fig:vorticity_up}). To further understand the detail of the instability cascade we distinguish between wavelengths equivalent to the incident wave (Fig. \ref{fig:vorticity_up}, blue volumes) and perturbations associated to higher horizontal modes (Fig. \ref{fig:vorticity_up}, red volumes). While the instability of the horizontal mode-1 structures is mostly stationary and occurs at both the lower and the higher jet edge we find transient regions of small Richardson numbers in the higher modes only above the jet. This is consistent with the growth of horizontal mode-2 structures due to the TRI as discussed above.

From these observations we may conclude that stationary structures with equal wave vector as the incident wave grow over time and eventually become unstable with respect to shear instability. What is more, higher modes generated through TRI may become unstable themselves over time posing an efficient pathway of the wave energy to small scales. The latter mechanism is observed at the upper jet edge only.

Here we would like to remind the reader that both the conditions for the shear and the static instabilities coincide with respect to the relative wave amplitude, $\alpha>1$, which is not fulfilled in the initial conditions. We thus interpret the observed shear instabilities as mechanisms occurring further downstream in the instability cascade.

\section{Conclusions}
\label{sec:conclusion}

In this study we have simulated a stationary internal gravity wave modulated by a sheared jet with amplitudes close to static instability in order to identify the cascade of both known and novel instability types. While some instability types like the classical modulational instability are excluded by choice many others show clear indications of occurrence. In particular, we find growing modes associated to the triadic resonant instability (TRI) mechanism as well as evidence of growing stationary structures with the same wave characteristic as the incident wave. While the transient TRI generated modes are dominant at the upper jet edge they are barely observed below. In contrast, the stationary structures occur at both the upper and lower edges of the jet albeit with an earlier onset and larger amplitude at the lower edge. These structures might be associated to the point spectrum modulational instability (PSMI) as proposed by \citet{Schlutow2020}, however it could not be identified unambiguously. Finally, the growing structures become subject to both shear and static instabilities ultimately leading to the breakdown of the incident wave and the transition to a turbulent regime.

We conclude with the remark that the breakdown of a modulated gravity wave is not associated to a single instability but a zoo of mechanisms occurring partly in parallel (competing for dominance) and in a cascade following one another or even breaking down the growing modes. Over all, this highlights that although many instability mechanisms are well known and understood many questions remain open. As an example the growth of stationary instabilities could not be uniquely associated to a specific mechanism in the present study. Thus we need more investigations to strengthen our understanding of the mechanics of gravity wave breaking.

\section*{Acknowledgements}

The authors like to thank the Center for Scientific Computing of the Goethe University Frankfurt. All calculations for this research were conducted on the provided Goethe-HLR Cluster. Moreover the authors thank Ulrich Achatz for his support and constructive feedback.

\section*{Funding}

This paper is a contribution to the project W01 (Gravity-wave parameterization for the atmosphere) and S02 (Improved Parameterizations and Numerics in Climate Models) of the Collaborative Research Centre TRR 181 "Energy Transfers in Atmosphere and Ocean" funded by the Deutsche Forschungsgemeinschaft (DFG, German Research Foundation) - Projektnummer 274762653.



%
\bibliographystyle{plainnat}
\bibliography{zotero.bib}

\end{document}